\documentclass[%
 reprint,
 amsmath,amssymb,
 aps,
]{revtex4-1}

\usepackage{graphicx}
\usepackage{dcolumn}
\usepackage{bm}

\begin{document}

\author{Sk Noor Alam}
 \email{noor1989phyalam@gmail.com}
\author{Subhasis Chattopadhyay}%
 \email{sub.chattopadhyay@gmail.com}
\affiliation{Variable Energy Cyclotron Centre, HBNI, 1/AF-Bidhannagar, Kolkata-700064, India}

\date{\today}%

\title{Effect of simulating parity-odd observables in high energy heavy ion collisions on balance functions of charged particles and elliptic flow of pions }

\begin{abstract}
At the early stage of heavy ion collisions, non-trivial topologies of the gauge fields can be created resulting in an imbalance of axial charge density and eventually separation of electric charges along the direction of the magnetic field produced in such collisions. This process is called the chiral magnetic effect (CME). In this work we implement such a charge separation at the partonic level in AMPT for Au+Au collisions at $\sqrt{s_{NN}}$ = 200 GeV to study its consequence on experimental observables. We present the effects on the pion elliptic flow ($v_2$) and the charged particle balance function (BF) for varying strengths of initial charge separation. We find that the shape of the balance function is sensitive to the increasing charge separation. $v_2$ of pion shows a strong decreasing trend at higher transverse momenta ($p_T$) with increasing charge separation. Charge balance functions show a peak at $\Delta\phi \sim 180 $ with charge separation implemented in the partonic level as expected for the parity violation. We have also calculated parity observable $\gamma$ in the form of BF's moments. $\gamma$ shows a decreasing trend with charge separation. It has a negative value for charge separation produced by flipping more than 30 $\%$ of quarks in the parton level. We also notice that $<\gamma>$ for the same charge correlation and the opposite charge correlation shows negative and positive values, respectively.
\end{abstract}

\maketitle

\section{Introduction}
\label{Section:Introduction}

In high energy heavy ion collisions, gluonic configuration such as sphalerons and instantons \cite{Ref:1st,Ref:2nd,Ref:3rd,Ref:4th,Ref:5th} can change the left-handed quarks to right-handed ones and vice versa through axial anomaly \cite{Ref:6th,Ref:9th,Ref:7th,Ref:8th}. The interactions between these gluonic configurations and the quarks break the parity (P) and charge conjugation parity (CP) symmetry due to this axial anomaly. This results in a net chirality due to P and CP breaking and generates an asymmetry between the number of left and right-handed fermions. The P and CP violations result in a separation between the positive and negative charges along the direction of magnetic field that is produced in heavy ion collisions into two hemispheres separated by the reaction plane (RP) \cite{Ref:10th} as determined by the direction of impact parameter and beam momentum. This phenomenon of QCD anomaly-driven charge separation is referred to as the chiral magnetic effect (CME). Several experimental measurements have been dedicated towards the search for the CME at RHIC and LHC\cite{Ref:11th,Ref:12th,Ref:13th,Ref:14th,Ref:15th,Ref:16th}.

It has been estimated that in high energy heavy-ion collisions, spectator protons produce a strong magnetic field $eB_y \approx m_{\pi}^2$ or $\sim$ $3.14\times10^{14}$ T \cite{Ref:4th}. A P- and CP-odd domain in the presence of a large magnetic field can generate
chirality by inducing $up-down$ asymmetry in the production of quarks and antiquarks. This asymmetry should be reflected in the final hadron production mainly of pions. An electric dipole moment pointed from the negative charge to the positive charge direction is created because of this charge separation.  \\
In this work, we have implemented this charge separation in a heavy-ion event generator known as A Multi-Phase Transport(AMPT) model via creating electric dipole moment at the partonic level.

We propose two observables widely used in heavy ion collisions i.e the balance function of charged particles and the elliptic flow of pions as the observables for the CME. \\ 

We first introduce the balance function. Balance function (BF) is a conditional probability of observing a charge with respect to another charge \cite{Ref:Pratt,Ref:Basspaper}. It is defined as 
\begin{equation}
B = \frac{1}{2} \Big[ \frac{\langle N_{(a,b)} \rangle}{\langle N_{a} \rangle} - \frac{\langle N_{(b,b)} \rangle}{\langle N_{b} \rangle} + \frac{\langle N_{(b,a)} \rangle}{\langle N_{b} \rangle} - \frac{\langle N_{(a,a)} \rangle}{\langle N_{a} \rangle} \Big], 
\label{Eq:bfDefinition}
\end{equation}

here $\frac{\langle N_{(a,b)} \rangle}{\langle N_{a} \rangle} $ is the conditional probability of observing a particle of type \textit{b} within a relative separation i.e $\Delta\eta=\mid\eta_{a}-\eta_{b}\mid$ or $\Delta\phi=\mid\phi_{a}-\phi_{b}\mid$ with respect to a particle type \textit{a}. Particle type \textit{a} and \textit{b} are used as trigger particle and associated particles respectively. $\langle N_{(a,b)} \rangle $ is the number of pairs that satisfies relative separation condition which is a function of relative separation and transverse momenta of trigger and associated particles i.e $\langle N_{(a,b)}(\Delta\eta,\Delta\phi,p_{\mathrm{T,trig}},p_{\mathrm{T,asso}}) \rangle $. In this paper we have studied the sensitivity of the BF structure with the varying fraction of charge separation. Here we have used the transverse momentum of the trigger particle higher than that of the associated particle. 
The balance function can be studied as a function of rapidity(y), pseudo-rapidity($\eta$) or azimuthal angle($\phi$)). We use acceptance cuts  in transverse momentum i.e $0.2 < p_{t} <2 $ GeV/c  and  in pseudorapidity i.e $\mid \eta \mid <1$.
\\

In Ref.\cite{Ref:obserCME}, it has been suggested that the gamma correlator i.e two particle correlation $\gamma$ is defined as $<\cos(\phi_{1}+\phi_{2}-2\psi_{RP})>$ where $\psi_{RP}$ is the reaction plane angle and $\phi_{1}$ , $\phi_{2} $ denote the azimuthal angles of the produced charged particles. $\gamma$ is sensitive to the CME effects.
The azimuthal distribution of produced particles with parity odd observables may have the following form
 \begin{equation}
        \frac{dN}{d\phi} \sim 1+\sum_{n=1}^{\infty}(2v_{n}\cos[ n(\phi-\Psi_{R}) ] + 2a_{n}\sin[ n(\phi-\Psi_{R}) ])
\label{Eq:parityOdd}
\end{equation} 
where $\phi$ is the azimuthal angle and $\Psi_{R}$ is the reaction plane angle. \textit{Sine} term represents the charge separation and the parameter $a_{n}$ describes the parity violation effect. We have calculated these parity violation terms in the form of balance function moments as discussed in ref.~\cite{Ref:37th}. The gist of this ref.~\cite{Ref:37th} has been discussed below.
We know $\gamma_{P}=\cos(\phi_{i}+\phi_{j}) =\cos(2\phi_{i})\cos(\Delta\phi)-\sin(2\phi_{i})\sin(\Delta\phi)$ where P stands for parity and $\Delta\phi=\phi_{j}-\phi_{i}$. $\gamma_{P}$ can be expressed when weighted with azimuthal distribution of particles as
\begin{equation}
\gamma_{P}= \langle C_{b} \cos(2\phi) \rangle - \langle S_{b} \sin(2\phi) \rangle 
\label{Eq:gammap}
\end{equation}

where

\begin{equation}
C_{b}= \frac{1}{Z_{b}} \int d\Delta\phi B(\Delta\phi) \cos(\Delta\phi) ,
\label{Eq:cb}
\end{equation}

\begin{equation}
S_{b}= \frac{1}{Z_{b}} \int d\Delta\phi B(\Delta\phi) \sin(\Delta\phi) , 
\label{Eq:sb}
\end{equation}

\begin{equation}
Z_{b}= \int d\Delta\phi B(\Delta\phi)  
\label{Eq:zb}
\end{equation}

$Z_{b}$ is integral of balance function used as normalization factor. We have used trigger and associated particles in the $\phi$ range of $-\pi$ to $\pi$. Balance function being a function of $\Delta\phi$ is therefore independent of $\phi$. $\langle C_{b} \cos(2\phi) \rangle$ and $\langle S_{b} \sin(2\phi) \rangle$ could therefore be written as $ C_{b} \langle \cos(2\phi) \rangle$ and $S_{b} \langle \sin(2\phi) \rangle$ respectively. \\

From Eq.\ref{Eq:parityOdd} one can get the n-th harmonic co-efficient defined as $v_{n}$ is $\langle\cos[ n(\phi-\Psi_{R}) ]\rangle$ where $\langle..\rangle$ denotes average over particles \cite{Ref:34th,Ref:35th}. The second Fourier coefficient $v_{2}$ called elliptic flow $\langle\cos(2\phi)\rangle$ is the quantity of our interest. In our simulation $\Psi_{R}$ is taken as 0 as per the implementation of AMPT model. In a non-central heavy ion collision, a pressure gradient in azimuthal angle is established because of the initial spatial anisotropy \cite{Ref:36th}. Due to this, pressure gradient along in-plane is higher than along the out-of-plane. So more particles are emitted in-plane than out-of-plane and it gives a positive elliptic flow coefficient. However, observed $v_{2}$ has also been explained in transport model as due to anisotropic escape of partons \cite{Ref:v21,Ref:v22}. Elliptic flow is sensitive to the CME effects because of out-of-plane charge separation and it shows a strong decreasing trend.

This paper is organized as follows, in section II we have described the AMPT model and in section III, we have described the method of charge separation that is implemented at the quark level. In section IV and V , we have discussed results and summary respectively. 

\section{A Multi-Phase Transport Model}
\label{Section:AMPT}

In this work, we have implemented charge separation at the partonic level in Au+Au collisions at $\sqrt{s_{NN}}$ = 200 GeV using the AMPT model. The AMPT model consists of different components with the heavy ion jet interaction generator (HIJING) to implement the initial conditions, Zhang's parton cascade (ZPC) for modelling the partonic scatterings, the Lund string fragmentation model or a quark coalescence model for hadronization, and a relativistic transport (ART) model for hadronic re-scattering\cite{Ref:18th,Ref:19th,Ref:20th}.
HIJING provides spatial and momentum distributions of the minijet partons and of the soft string excitations \cite{Ref:21th,Ref:22th}. The cascading of partons are carried out using the ZPC model \cite{Ref:23th}. Partonic cross sections between 1 and 3 mb have been used for flow like studies using this model. However, in Ref \cite{Ref:32th} in which CME effect has been simulated using AMPT, parton cross-section of 10 mb has been used to explain STAR data. We have, therefore, used 10 mb partonic cross-section for this work. AMPT model has two versions , one is the Default AMPT and the other is the string melting(SM) version. In the default AMPT model, partons are combined with their parent strings when they stop interacting and the resulting strings are converted to hadrons using the Lund string fragmentation model \cite{Ref:24th,Ref:25th,Ref:26th}. In the AMPT with string melting \cite{Ref:27th,Ref:28th,Ref:29th}, a quark coalescence model is used instead to combine partons into hadrons. In the string melting mechanism, all excited strings that are not from the projectile or the target nucleons or without any interactions are converted to partons according to the flavor and spin structures of their valence quarks. Subsequently the dynamics of the hadronic matter is described by a hadronic cascade, which is based on the ART model \cite{Ref:30th,Ref:31th}. In the ART model, charges are not conserved. We have used NTMAX = 3 for minimising rescattering among hadrons. If one excludes hadron evolution in string melting model, main contribution of evolution are carried by parton cascade.

\begin{figure}[htb]
\centering
\includegraphics[width=.50\textwidth]{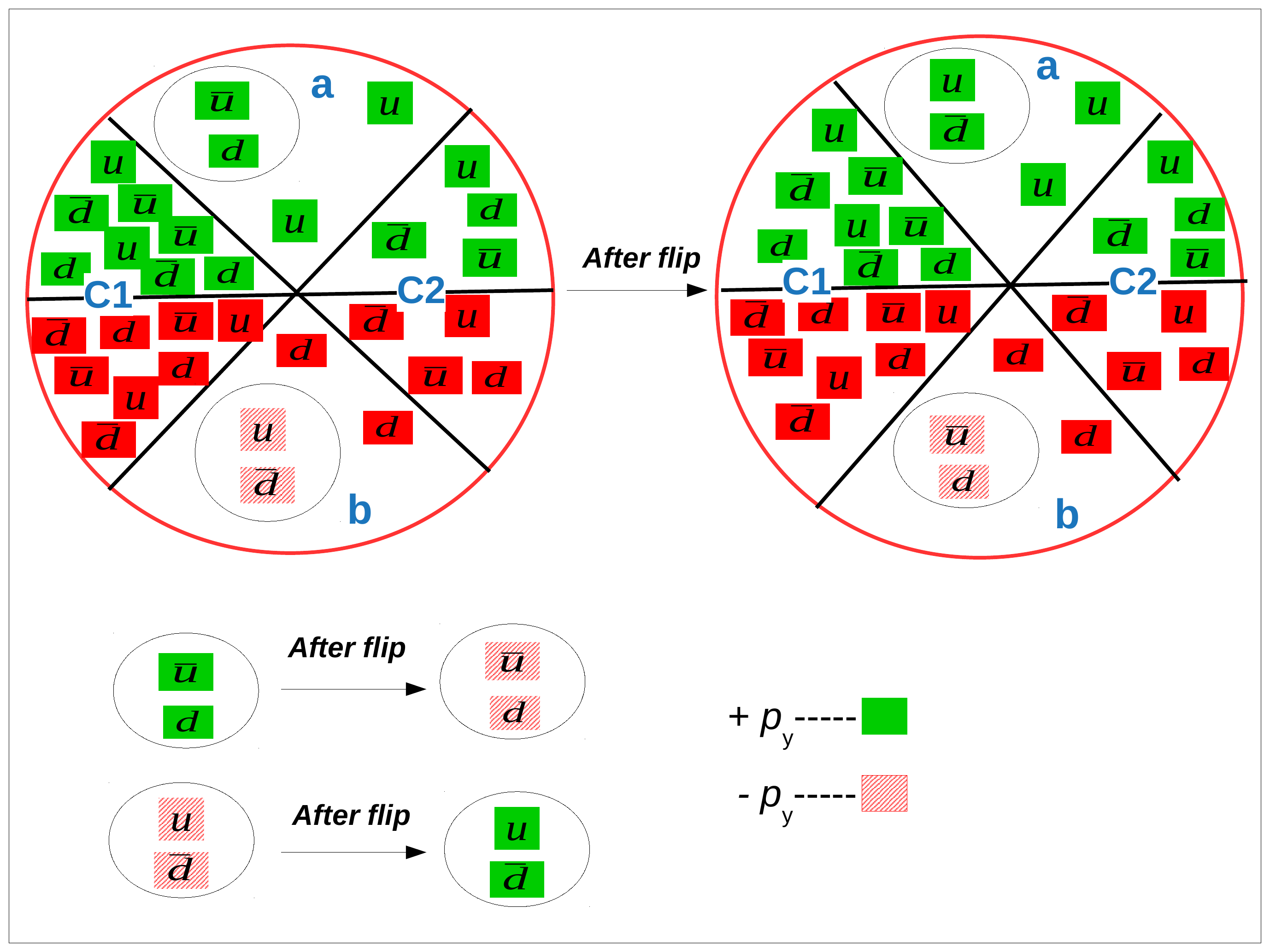}
\caption{(Color online) Charge Separation mechanism shown schematically}
\label{fig:FlipFigure}
\end{figure}

\section{Procedure of generating Charge Separation}
\label{Section:Procedure of generating CME}

The manifestation of the chiral magnetic effect is seen by a charge separation along the direction of the magnetic field. The charge separation is a result of P and CP odd domains. According to theory, in non-central heavy ion collisions, spectator protons create a magnetic field perpendicular to the reaction plane. We thus introduce a charge separation perpendicular to the reaction plane. Fig.~\ref{fig:FlipFigure} shows schematically the method of charge separation mechanism we have implemented in AMPT-SM model. 

In AMPT, RP angle is at 0$^{\circ}$. To have a direction of charge separation perpenidcular to RP, we first choose u, $\bar{\textit{u}}$, d and $\bar{\textit{d}}$ which have azimuthal angle between $\mid 1.0472^c \mid$ to $\mid 2.0944^c\mid$ i.e lying in regions $\bf{a}$ and $\bf{b}$ in Fig.~\ref{fig:FlipFigure}. $p_{y}$ momenta of the quarks in those selected regions are then modified in such a way that it results in a net positive charge in the upward direction and a net negative charge in the downward direction. Please note that quarks of other regions remain unchanged. The main purpose of selecting quarks from these two regions .i.e region $\bf{a}$ and $\bf{b}$ is to simulate a scenario of the CME where charge separation is created perpendicular to the RP and in-plane quarks remain unaffected.

 To achieve this, we replace a fraction of total number of upward going negatively charged quarks with downward going positively charged quarks and vice versa. In practice, $-p_{y}$ of a positively charged u quark and $+p_{y}$ of a negatively charged $\bar{\textit{u}}$ quark are flipped to each other making positively charged quark upgoing and negatively charged quark downgoing. Similarly flipping takes place between the $+p_{y}$ of a negatively charged d quark with the $-p_{y}$ of a positively charged $\bar{\textit{d}}$ quark. This charge separation method was used in Ref.\cite{Ref:32th}. As shown in the Fig.~\ref{fig:FlipFigure}, before flipping, each of the regions marked with $\bf{a}$ and $\bf{b}$ lying perpendicular to the reaction plane is with net-charges of $\frac{1}{3} e $.  Now after flipping the corresponding regions are with charges of $\frac{7}{3} e$ and$ -\frac{5}{3} e $ respectively thereby generating a charge separation perpendicular to the reaction plane.

\begin{figure}[htb]
\centering
\includegraphics[width=.50\textwidth]{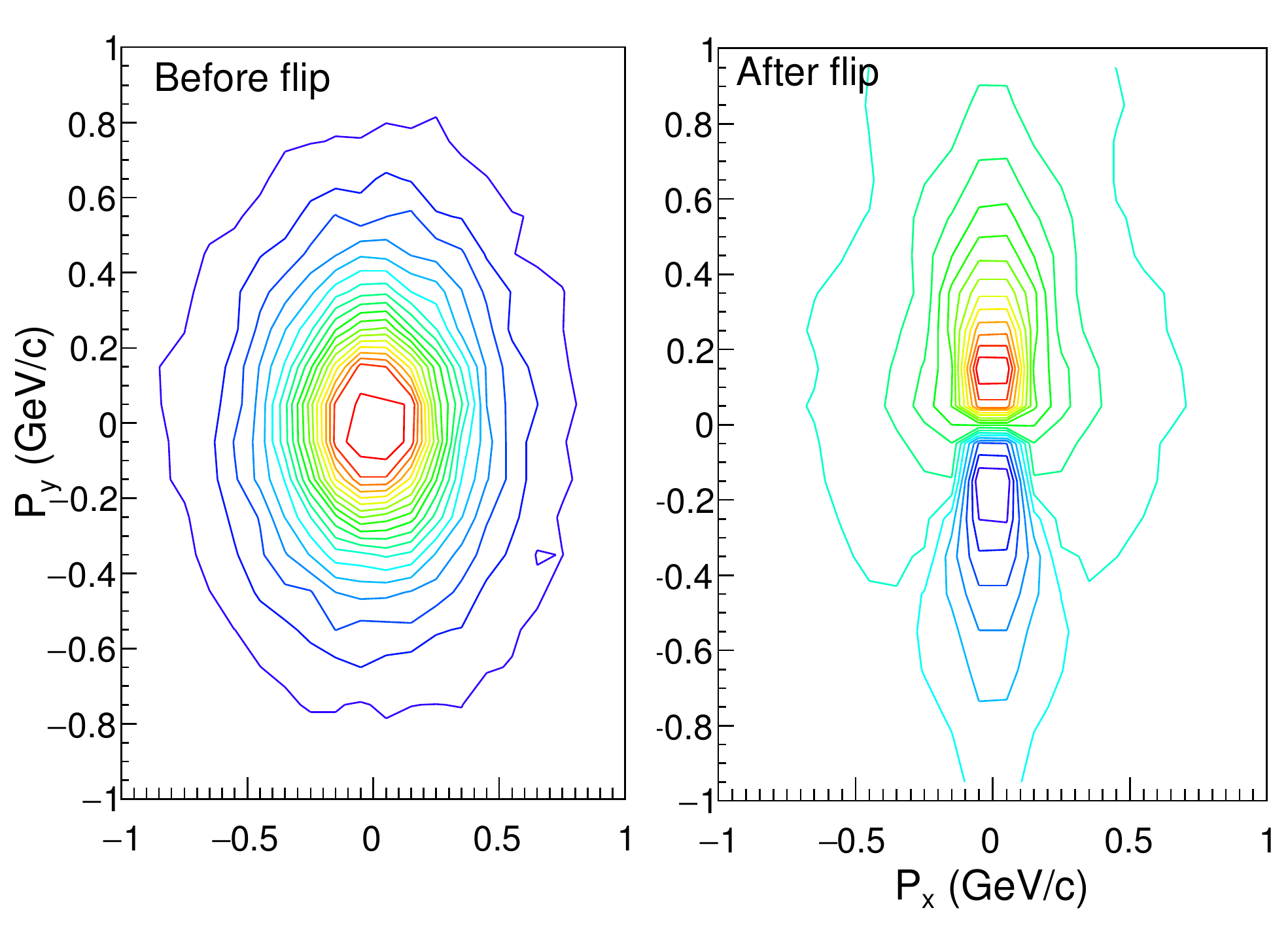}
\caption{(Color online) Net electric charge distributions on the transverse plane before and after flipping with a $20\%$ flipping fraction.}
\label{fig:1}
\end{figure}

After the implementation of flipping at the partonic level, the evolution of the system follows. The fraction (\textit{f}) of the total number of quarks that have been flipped is taken as an input parameter. We have calculated multiplicities of $\bar{\textit{u}}$ and \textit{d} quarks separately in the region $\bf{a}$ i.e $M_{\bar{\textit{u}}}^{\bf{a}}$ and $M_{\textit{d}}^{\bf{a}}$ respectively. Similarly multiplicities $M_{\textit{u}}^{\bf{b}}$ $\&$ $M_{\bar{\textit{d}}}^{\bf{b}}$ of \textit{u} and $\bar{\textit{d}}$ quarks respectively in region $\bf{b}$ have been obtained. $M_{small}^{\bar{\textit{u}},\textit{u}}$ = min ( $M_{\bar{\textit{u}}}^{\bf{a}} , M_{\textit{u}}^{\bf{b}}$ ) $\&$ $M_{small}^{\bar{\textit{d}},\textit{d}}$ = min ( $M_{\bar{\textit{d}}}^{\bf{b}} , M_{\textit{d}}^{\bf{a}}$ ). We then calculate $\textit{f}\times M_{small}^{\bar{\textit{u}},\textit{u}}$ for every event and this number is the number of ( $\textit{u} , \bar{\textit{u}}$ ) quarks to be flipped by exchange of $p_{y}$ momenta. Similar procedure has been followed for \textit{d} and $\bar{\textit{d}}$. In this work, $\textit{f}$ = 0, 0.1, 0.2, 0.3, 0.4, 0.5, and 0.6 have been used. The AMPT-SM has been used in which the quarks are hadronized by coalescence method as discussed earlier. The observables discussed in section I have been studied for the finally produced hadrons. In high-energy heavy-ion collisions, these observables might be studied with centrality as the magnitude of the magnetic field created in such collisions depends on centrality. In the present study, different magnitudes of charge separation as given by \textit{f} represent different magnitudes of the produced magnetic field and can be compared with collisions of various centralities. 

\begin{figure}[htb]
\centering
\includegraphics[width=0.50\textwidth]{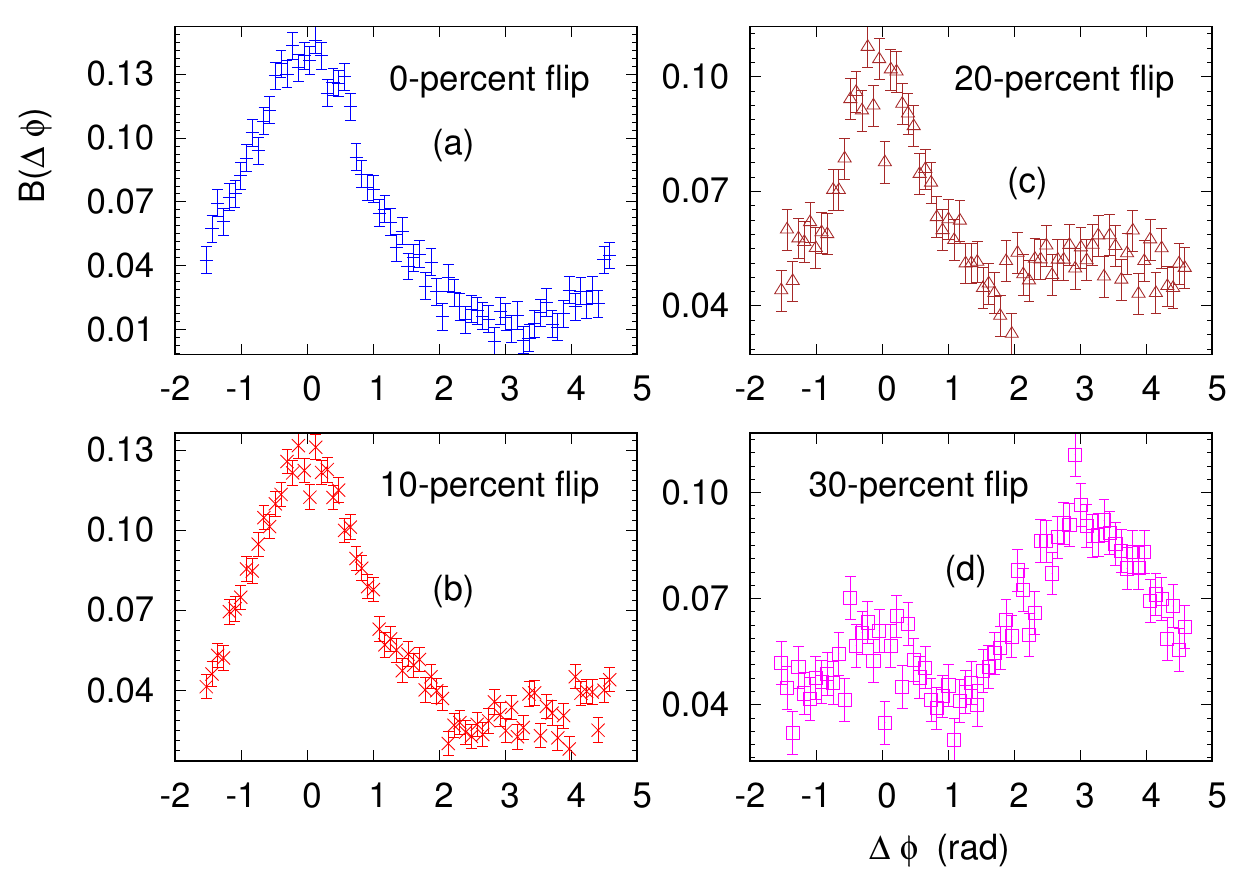}
\includegraphics[width=0.50\textwidth]{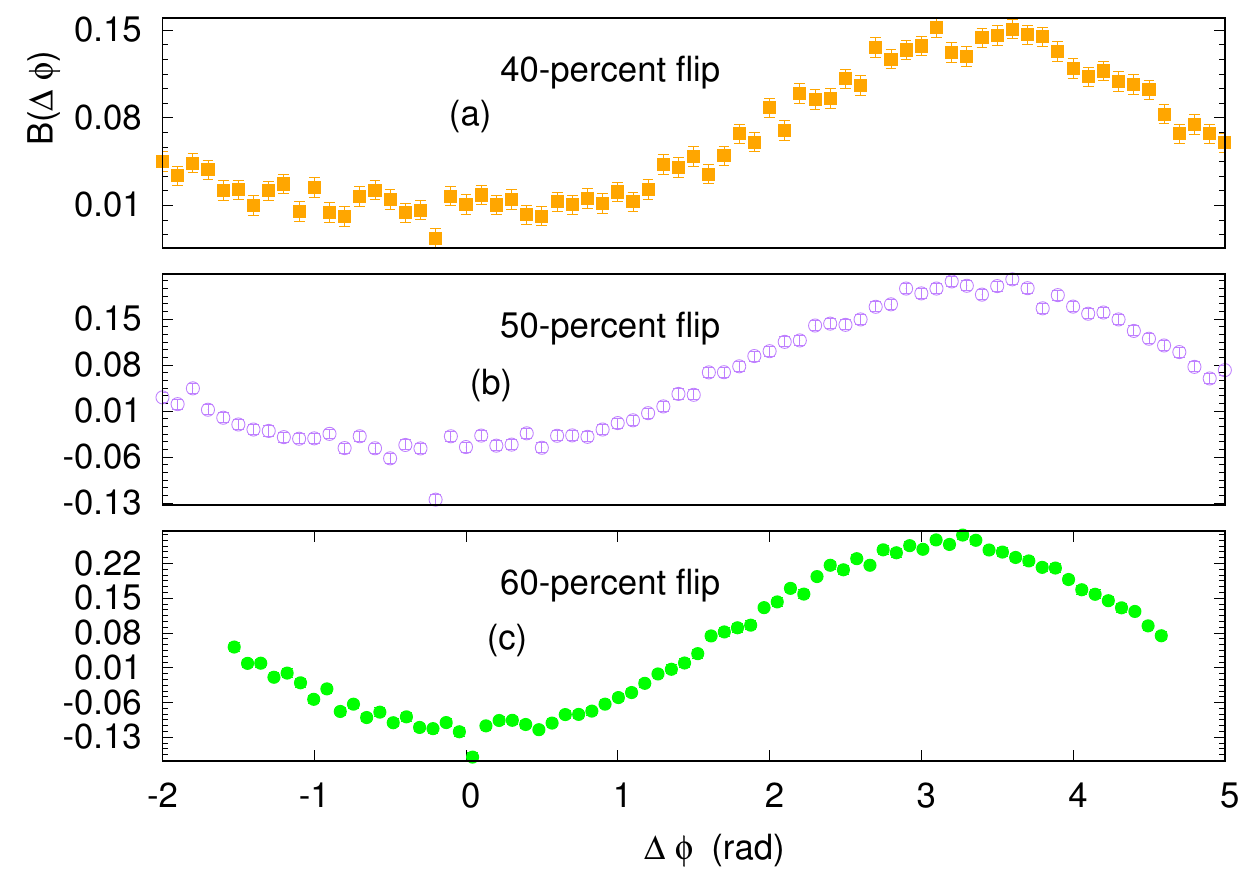}
\caption{(Color online) BF for Au+Au minimum bias at $\sqrt{s_{NN}}$ = 200 GeV with different flipping fractions from 0 to 60 $\%$}
\label{fig:4}
\end{figure}


\section{Results}
\label{Section:Results}

Fig.~\ref{fig:1} shows the net electric charge distributions on the transverse plane before and after flipping of quarks in Au+Au collisions at $\sqrt{s_{NN}}$ = 200 GeV. The contours indicate the net charge density profile. One can see that the distribution is symmetric when there is no charge separation. There is a net electric charge distribution on the transverse plane after flipping 20 $\%$ of quarks. It is clearly observed that an out-of-plane charge separation is generated after introduction of charge separation effects in the AMPT model.

\begin{figure}[htb]
\centering
\includegraphics[width=.50\textwidth]{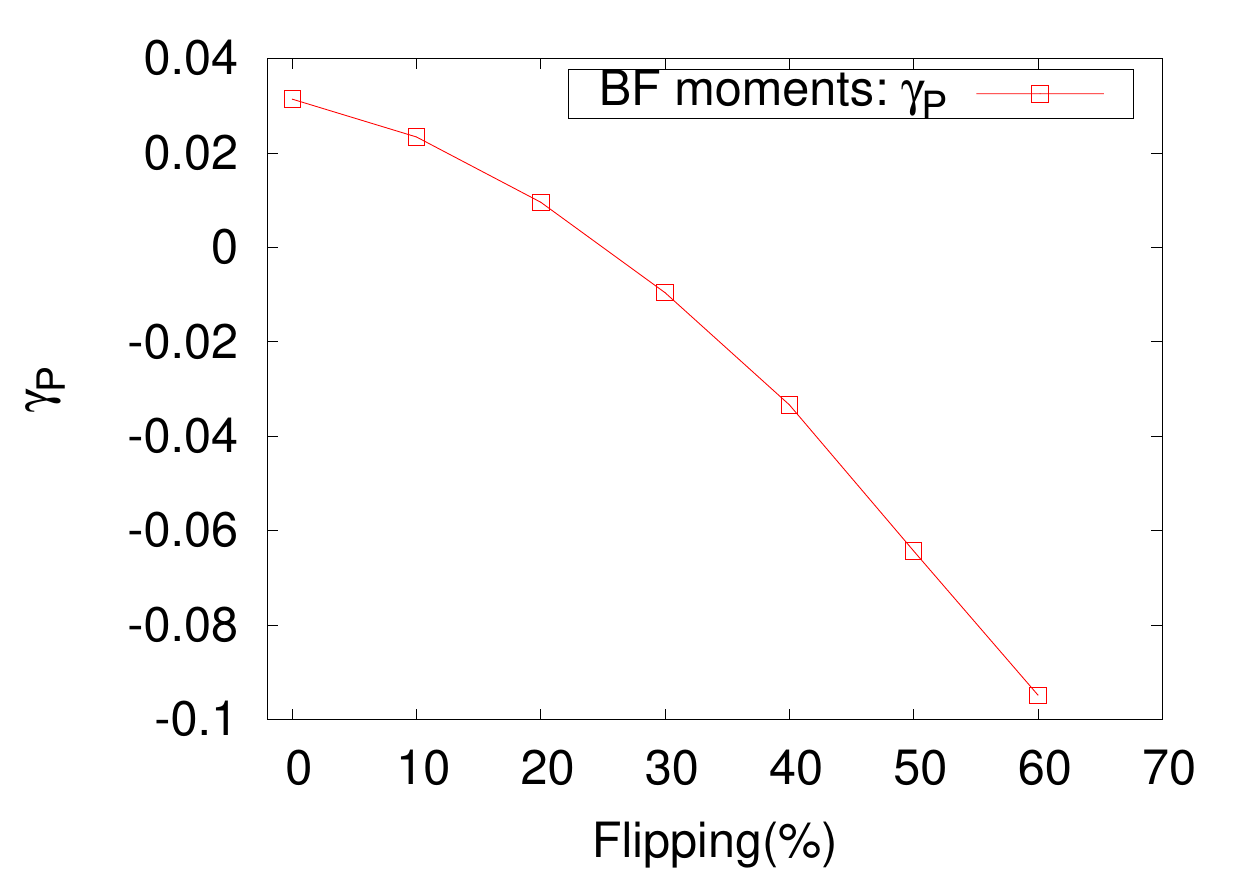}
\caption{(Color online) $\gamma$-correlator in the form of BF's moment with different flipping fractions}
\label{fig:7}
\end{figure}

\begin{figure}[htb]
\centering
\includegraphics[width=.50\textwidth]{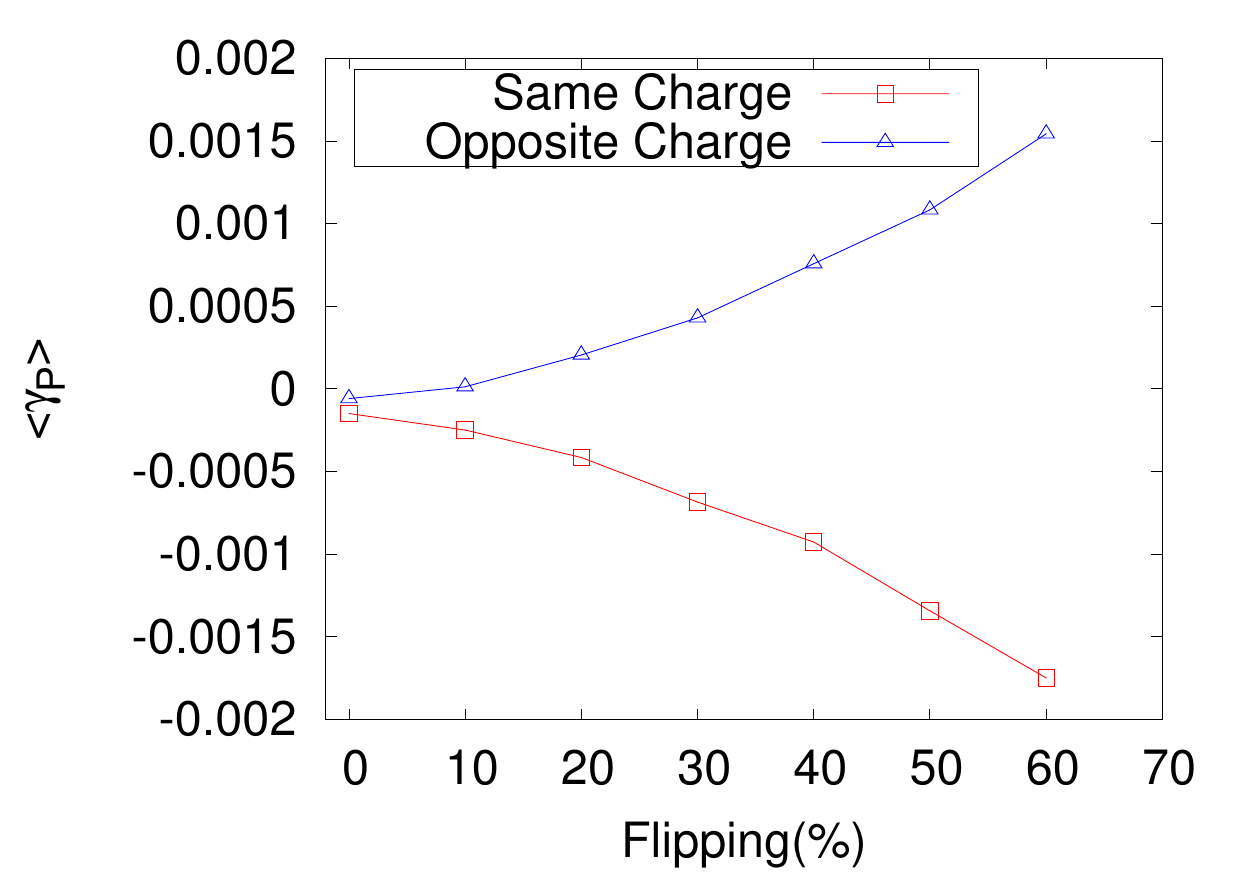}
\caption{(Color online) $<\gamma_{P}>$ with different flipping fractions}
\label{fig:8}
\end{figure}

Fig.~\ref{fig:4} show the charged particle balance function at different flipping fractions. It is clearly seen that the shape of the balance function evolves with the flipping fraction. While without flipping, it shows a peak$\sim$ $0^c$, the peak shifts towards $\pi^c$ when flipping for charge separation is 30 $\%$ or greater. For parity violation, balance function should have a peak at $\Delta\phi \sim \pi^c$ \cite{Ref:37th}.  We have studied the effect of the widely used observable i.e $\gamma$-correlator on BF. $\gamma_{P}$ correlator in the form of BF's moment as defined in Eq. \ref{Eq:gammap} is shown in Fig.~\ref{fig:7}. We have also shown $\gamma_{P}$  averaged with $N_{part}$ (number of participants) distribution as defined in Eq. \ref{Eq:averageGamm} in Fig.~\ref{fig:8}.

\begin{equation}
\langle \gamma_{P}(N_{part}) \rangle= \frac{\int d(N_{part})\gamma_{P}(N_{part}) N_{part}} {\int d(N_{part}) N_{part}}
\label{Eq:averageGamm}
\end{equation} \\ where $\int d(N_{part}) N_{part}$ is the $N_{part}$ distribution. In this work, we have used impact parameter range from 0 to 12 fm and $N_{part}$ in the range of 21 to 392. \\

Fig.~\ref{fig:7} shows that the parity odd observable in form of BF moments becomes negative when flipping fraction is $\sim$ 30 $\%$ or higher. This signifies the presence of more balancing pairs in out-of-plane relative to that of in-plane direction. It may be noted that this effect is due to the $2^{nd}$ term in Eq.~\ref{Eq:gammap} which arises due to the CME effect. \\

Fig.~\ref{fig:8} shows that $<\gamma_{P}>$ with same charges and opposite charges have negative and positive values respectively. This figure indicates that gamma correlator has larger magnitudes with higher flipping fraction for both same and opposite charge correlation. $\gamma$-correlator used in \cite{Ref:12th} as an observable also shows similar trend thereby showing that the CME effect implemented in AMPT model is reasonable. It should be mentioned that Fig.~\ref{fig:8} in this work and plot in Ref.\cite{Ref:32th} might look similar. However, in this work, in Fig.~\ref{fig:8}, we have plotted the gamma correlator averaged with $N_{part}$ vs flipping fractions. In Ref.\cite{Ref:32th}, gamma correlator is plotted with different centralities. As the event with different flipping fractions might be related with centrality, a connections may be drawn between two cases. 

\begin{figure}[htb]
\centering
\includegraphics[width=.50\textwidth]{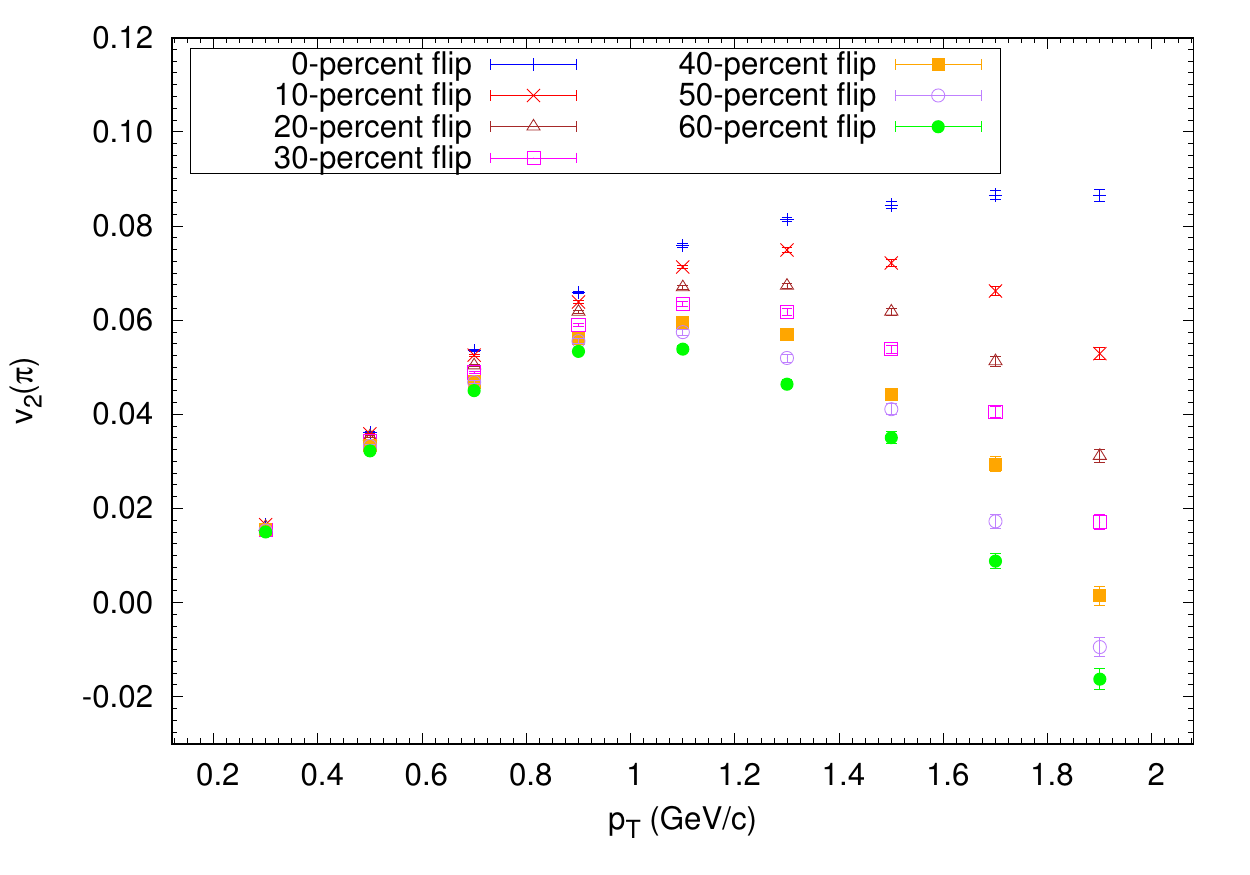}
\caption{(Color online)Elliptic flow of pions for different flipping fractions and no flip }
\label{fig:6}
\end{figure}

Fig.~\ref{fig:6} shows the elliptic flow($v_2$) of pions as a function of $p_{T}$ for different flipping fractions in the initial partonic state of the collisions. It is observed that the elliptic flow of pion increases upto $p_{T}$ $\sim$ 1.1 GeV/c and then decreases at higher $p_{T}$. There is an increase in out-of plane particle production , so $v_{2}$ shows a decreasing trend for higher flipping fractions. 


\section{Summary}
\label{Section:Summary}

In this study, momenta of initial partons of AMPT generator have been flipped to generate an out-of-plane charge separation in Au+Au collisions at $\sqrt{s_{NN}}$ = 200 GeV. The fraction of a type of quark(\textit{u},$\bar{\textit{u}}$,\textit{d},$\bar{\textit{d}}$) has been taken as a variable. This charge separation represents the effect of parity-odd observable in heavy ion collisions where magnetic fields are generated. We have studied the effect of this charge separation on two widely used observables i.e charge particle BF and elliptic flow of pions. $\gamma$-correlator has also been used for comparison. The observables are chosen in such a way that they characterize the effect of net-charge and their distribution on azimuthal plane. Different fractions represent varying centrality in such collisions. 
In this study, with varying fraction of flipping, both the BF and $v_2$ show significant sensitivity with the peak of the BF shifting from $\Delta\phi=0$ towards $\Delta\phi=\pi$ with increasing flipping fraction and $v_2$ of pions decreases at higher $p_T$. The reduction in $v_2$ with respect to no-flipping scenario depends on the flipping fraction. The gamma correlator in form of BF moment with different flipping fraction shows a decreasing trend. We also notice that $<\gamma_{P}>$ for same charge correlation and opposite charge correlation have opposite values and varies with charge separation. Experimentaly, the STAR has an upper limit for the value of gamma correlator of the order of $10^-3$. We have observed that the gamma correlator of $\approx 10^-3$ corresponds to flipping fractions range of 0 to 60$\%$. We hereby propose to look at both the observables i.e BF and elliptic flow together for making an unambiguous conclusion on the generation of parity-odd effects in high energy heavy ion collisions.  

\section*{Acknowledgements}
\label{Section:Acknowledgements}

This work has used resource of grid computing facility at Variable Energy Cyclotron Centre, Kolkata. We are grateful to Prithwish Tribedy for helpful discussions.

\end{document}